%
%
%
%
%
\input mtexsis             

\let\chi=\oldchi
\let\zeta=\oldzeta
\preprint                  
\bookpagenumbers

%
%
%


\def\Tbls#1{Tables~\use{Tb.#1}}
\def\Tblp#1{\use{Tb.#1}}


%
\begingroup   
   \ATunlock
   \offparens                                   
   \gdef\j@urnal#1;#2,#3(#4){
      {#4,} {\it#1\/} {\vol{#2},} #3\relax      
   \egroup}
   \ATlock
\endgroup
\def\vol#1{{\bf#1}}   




\newif\ifthesis   

%
\newdimen\figsize
\newdimen\smallfigsize
\newdimen\bigfigsize
\figsize=0.707107\colwidth  
\smallfigsize=8cm
\bigfigsize=17cm
%


\def\reps{{\rm I\kern-.19em P}}





\def\x{\;\pmb{\mit x}}         


\def\zn{{{}_{0n}}}	

\hyphenation{axi-sym-metric}






\def\eg{{\it e.g.\/}}

\def\abs#1{\left|#1\right|}
\def\inprod#1{\left\langle#1\right\rangle}

\def\di#1#2{{\partial #1\over \partial #2}}


\def\sech{\,{\rm sech}}
%
\def\pmb#1{\setbox0=\hbox{$#1$}%
  \kern-0.25em\copy0\kern-\wd0
  \kern.05em\copy0\kern-\wd0
  \kern-0.025em\raise.0433em\box0}
%
\def\reals{{\rm I\kern-.19em R}}
\def\naturals{{\rm I\kern-.19em N}}
%
\long\def\comment#1{ $\langle\!\langle${\tenpoint\bf#1}$\rangle\!\rangle$ }
\long\def\dcomment#1{ $\langle\!\langle${\sl#1}$\rangle\!\rangle$ }
\long\def\fcomment#1{ $\langle\!\langle${\it#1}$\rangle\!\rangle$ }
\long\def\gobble#1{}
\def\nocomments{%
  \def\comment{\gobble}\def\dcomment{\gobble}\def\fcomment{\gobble}}

%
\newcount\footnotenum
\footnotenum=1
\long\def\FNT#1{\Footnote{{\the\footnotenum}}{#1}\global\advance\footnotenum by 1}

%
\def\fref#1{\advance\footnotenum by -#1 \the\footnotenum
     \advance\footnotenum by #1 }
%
\def\papertitle#1{{\rm``#1''}}
%


\def\AAp{Astron.\ Astrophys.}

\def\ApJ{Astrophys.~J.}

\def\ApSS{Astrophys.\ and Space Sci.}

\def\MNRAS{Mon.\ Not.\ R.\ astr.\ Soc.}


\def\AP{Academic Press}
\def\CUP{Cambridge University Press}

\def\PUP{Princeton University Press}


\thesisfalse               
\nocomments
\overfullrule=0pt
%

\def\scriptL{{I\!\!L}}
\def\Rden{\rho_{{}_R}}
\def\zden{\rho_{{}_z}}
\def\Sqrt#1{\left(#1\right)^{1/2}}
\def\erfc{\,{\rm erfc}}
\def\csch{\,{\rm csch}}
\def\R{{\cal R}}
\def\Z{{\cal Z}}
\def\zn{z_{{}_0}}
\def\ftwh{{\tilde w}_h}
\def\ftzden{\tilde\zden}
\def\integers{Z\!\!\!Z}
\def\hyg{{}_2F_1}
\def\length{L}

%
%

\pubdate{18 January 1996}         
\pubcode{Version 1.1}             

\bigskip
\bigskip
\titlepage

\title
Potential-Density Basis Sets for Galactic Disks
\endtitle

\author
David J.D.~Earn
Racah Institute of Physics, The Hebrew University, Jerusalem 91904, Israel
\smallskip
E-mail: earn@astro.huji.ac.il
\endauthor

\abstract

A class of complete potential-density basis sets in cylindrical
$(R,\phi,z)$ coordinates is presented.  This class is suitable for
stability studies of galactic disks in three dimensions and includes
basis sets tailored for disks with vertical density profiles that are
exponential ($e^{-|z|/\zn}$), Gaussian ($e^{-(z/\zn)^2}$) or locally
isothermal ($\sech^2(z/\zn)\,$).  
The basis sets are non-discrete and non-biorthogonal; however, the
extra numerical computations required (compared with discrete
biorthogonal sets) are explained and constitute a small overhead.
The method of construction (and
proof of completeness) is simple and can be used to construct basis
sets for other density distributions that are best described in
circular or elliptic cylindrical coordinates.  When combined with a
basis set designed for spheroidal systems, the basis sets presented
here can be used to study the stability of realistic disks embedded in
massive halos.

\endabstract

\bigskip
{
\noindent{\it Subject headings}: galaxies: kinematics and dynamics 
--- galaxies: structure --- instabilities \hfill\break
\null\hskip3.3cm --- methods: numerical --- celestial mechanics: stellar dynamics
}

\bigskip
\bigskip
\bigskip
\toappear{The Astrophysical Journal}

\endtitlepage


\nopagenumbers
\null\vfill\supereject	
\pagenumbers
\pageno=1

\expandafter \gdef \csname @Eq.Poisson@\endcsname {1.1}
\expandafter \gdef \csname @Eq.expand@\endcsname {1.2}
\expandafter \gdef \csname @Eq.expand;a@\endcsname {1.2a}
\expandafter \gdef \csname @Eq.expand@\endcsname {1.2}
\expandafter \gdef \csname @Eq.expand;b@\endcsname {1.2b}
\expandafter \gdef \csname @Eq.densep@\endcsname {1.3}
\expandafter \gdef \csname @Eq.exp@\endcsname {1.4}
\expandafter \gdef \csname @Eq.eval@\endcsname {2.1}
\expandafter \gdef \csname @Eq.sep@\endcsname {2.2}
\expandafter \gdef \csname @Eq.ode@\endcsname {2.3}
\expandafter \gdef \csname @Eq.ode;a@\endcsname {2.3a}
\expandafter \gdef \csname @Eq.ode@\endcsname {2.3}
\expandafter \gdef \csname @Eq.ode;b@\endcsname {2.3b}
\expandafter \gdef \csname @Eq.ode@\endcsname {2.3}
\expandafter \gdef \csname @Eq.ode;c@\endcsname {2.3c}
\expandafter \gdef \csname @Eq.odesoln@\endcsname {2.4}
\expandafter \gdef \csname @Eq.odesoln;a@\endcsname {2.4a}
\expandafter \gdef \csname @Eq.odesoln@\endcsname {2.4}
\expandafter \gdef \csname @Eq.odesoln;b@\endcsname {2.4b}
\expandafter \gdef \csname @Eq.odesoln@\endcsname {2.4}
\expandafter \gdef \csname @Eq.odesoln;c@\endcsname {2.4c}
\expandafter \gdef \csname @Eq.pair@\endcsname {2.5}
\expandafter \gdef \csname @Eq.pair;a@\endcsname {2.5a}
\expandafter \gdef \csname @Eq.pair@\endcsname {2.5}
\expandafter \gdef \csname @Eq.pair;b@\endcsname {2.5b}
\expandafter \gdef \csname @Eq.idea@\endcsname {3.1}
\expandafter \gdef \csname @Eq.idea;a@\endcsname {3.1a}
\expandafter \gdef \csname @Eq.idea@\endcsname {3.1}
\expandafter \gdef \csname @Eq.idea;b@\endcsname {3.1b}
\expandafter \gdef \csname @Eq.zode@\endcsname {3.2}
\expandafter \gdef \csname @Tb.nok@\endcsname {3.1}
\expandafter \gdef \csname @Eq.*@\endcsname {3.3}
\expandafter \gdef \csname @Tb.withk@\endcsname {3.2}
\expandafter \gdef \csname @Eq.newpair@\endcsname {4.1}
\expandafter \gdef \csname @Eq.newpair;a@\endcsname {4.1a}
\expandafter \gdef \csname @Eq.newpair@\endcsname {4.1}
\expandafter \gdef \csname @Eq.newpair;b@\endcsname {4.1b}
\expandafter \gdef \csname @Eq.Fredholm@\endcsname {4.2}
\expandafter \gdef \csname @Eq.*@\endcsname {4.3}
\expandafter \gdef \csname @Eq.wh@\endcsname {4.4}
\expandafter \gdef \csname @Tb.ft@\endcsname {4.1}
\expandafter \gdef \csname @Eq.innerprod@\endcsname {4.5}
\expandafter \gdef \csname @Eq.*@\endcsname {4.6}
\expandafter \gdef \csname @Eq.Mij@\endcsname {4.7}
\expandafter \gdef \csname @Eq.*@\endcsname {4.8}
\expandafter \gdef \csname @Eq.PD@\endcsname {4.9}
\expandafter \gdef \csname @Eq.PD;a@\endcsname {4.9a}
\expandafter \gdef \csname @Eq.PD@\endcsname {4.9}
\expandafter \gdef \csname @Eq.PD;b@\endcsname {4.9b}
\expandafter \gdef \csname @Eq.*@\endcsname {4.10}
\expandafter \gdef \csname @Eq.biorthonormal@\endcsname {4.11}
\expandafter \gdef \csname @Eq.*@\endcsname {4.12}
\expandafter \gdef \csname @Eq.*@\endcsname {4.13}
\expandafter \gdef \csname @Eq.*@\endcsname {4.14}
\expandafter \gdef \csname @Eq.*@\endcsname {4.15}
\expandafter \gdef \csname @Eq.ckmh@\endcsname {4.16}
\expandafter \gdef \csname @Eq.inteq@\endcsname {4.17}
\expandafter \gdef \csname @Eq.int2sum@\endcsname {5.1}
\expandafter \gdef \csname @Eq.int2sum;a@\endcsname {5.1a}
\expandafter \gdef \csname @Eq.int2sum@\endcsname {5.1}
\expandafter \gdef \csname @Eq.int2sum;b@\endcsname {5.1b}
\expandafter \gdef \csname @Eq.*@\endcsname {5.2}
\expandafter \gdef \csname @Eq.mateq@\endcsname {5.3}
\expandafter \gdef \csname @Eq.mateq;a@\endcsname {5.3a}
\expandafter \gdef \csname @Eq.mateq@\endcsname {5.3}
\expandafter \gdef \csname @Eq.mateq;b@\endcsname {5.3b}
\expandafter \gdef \csname @Eq.*@\endcsname {5.4}
\expandafter \gdef \csname @Eq.dis1@\endcsname {5.5}
\expandafter \gdef \csname @Eq.dis2@\endcsname {5.6}
\expandafter \gdef \csname @Eq.dis3@\endcsname {5.7}

\section{Introduction}

The first step in building a galaxy model is to find or approximate
the potential $\psi$ generated by a model mass density $\rho$.
Poisson's equation,
$$
\nabla^2\psi=4\pi G\rho \,,
\EQN Poisson
$$
can be solved using basis sets of potential-density
pairs $\{(\psi_j,\rho_j) : \nabla^2\psi_j=4\pi G\rho_j\}$.  A given density
$\rho$ is expanded in the basis density functions,
$$
\rho = \sum_{j} c_j \rho_j \,;
\EQN expand;a
$$
since \Eq{Poisson} is linear, the corresponding potential is
$$
\psi = \sum_{j} c_j \psi_j \,,
\EQN expand;b
$$
with the same coefficients $c_j$.  An approximate solution is obtained
using finitely many terms.

In this way we can approximate the potentials (and force fields) of
mass distributions that are not amenable to an exact analytical
solution of \Eq{Poisson}.  Moreover, basis expansions are fundamental
to semi-analytical normal mode analyses of stellar systems (Kalnajs
1977) and to a powerful $N$-body simulation technique (Clutton-Brock
1972) that is ideally suited to stability studies (Earn \& Sellwood
1995).  The first step required before implementing these methods is
to find a basis set that is well-suited to the model of interest.

The eigenfunctions of the Laplacian operator $\grad^2$ always form a
complete biorthogonal basis set (\eg, Courant \& Hilbert 1953, \S6.3;
Arfken 1985, \S9.4) but they do not always converge sufficiently fast
to typical potentials relevant for galactic dynamics.  It is essential
that a given model and its normal modes can be represented with a
modest number of basis functions; otherwise, accurate expansions
become prohibitively expensive.

A variety of useful sets have been derived for flat disks
(Clutton-Brock 1972; Kalnajs 1976; Qian 1992, 1993) and for spheroidal
systems (Clutton-Brock 1973; Hernquist \& Ostriker 1992; Saha 1993;
Syer 1995; Robijn \& Earn 1996; Zhao 1996).  However, the literature
has apparently been void of sets that are well-suited to disks of
finite thickness.  Such basis sets are needed for studies of realistic
three-dimensional (3D) disk galaxy models.

There is no evidence that the vertical structure of galactic disks varies
significantly with radius.  The goal of the present paper is therefore
to find basis sets suitable for studies of 3D disks with mass
densities of the separable form
$$
\rho(R,z) = \Rden(R)\zden(z) \,.
\EQN densep
$$
Observations indicate that the distribution of luminous mass in disk
galaxies is well-approximated by \Ep{densep} with
$$
\Rden(R) = {M\over 2\pi a^2 }\,e^{-R/a}\,, 
\qquad \zden(z) = {1\over 2b}\sech^2(z/b) \,,
\EQN exp
$$
or with $\zden(z)\propto e^{-\abs{z}/b}$ or some other power of
$\sech(z/b)$ (\cf\ Freeman 1970; van der Kruit 1988).  The vertical
density profile $\zden(z)\propto\sech^2(z/b)$ is favoured
theoretically because it results if we demand that the disk be locally
isothermal (Spitzer 1942; Binney \& Tremaine 1987, problem 4-25).  In
\Ep{exp}, $M$ is the total mass and $a$ and $b$ are scale lengths.

New basis sets for 3D disks are derived in this paper (some of the
main results were summarized briefly by Earn 1995) and a simple proof
of completeness is also provided.  A derivation of the Laplacian
eigenfunctions is reviewed below; the present approach is essentially
a modification of this procedure, together with a simple trick.
Different methods for constructing basis sets for 3D disks are
discussed by Robijn \& Earn (1996).

\section{The Standard Basis}

As a first step, we derive the set of eigenfunctions for the Laplacian
operator in cylindrical $(R,\phi,z)$ coordinates.  Thus we seek
solutions of
$$
\grad^2 \psi
    = {1\over R} \di{}{R} ( R \di{\psi}{R} )
      + {1\over R^2} \di{^2\psi}{\phi^2}
      + \di{^2\psi}{z^2}
    = \lambda\psi \;.
\EQN eval
$$
This will give potential-density pairs that are proportional
($\rho = {\lambda\over4\pi G}\psi$).  Separating variables,
$$
\psi(R,\phi,z) = \R(R) \Phi(\phi) \Z(z) \;,
\EQN sep
$$
we obtain from \Eq{eval} the three ordinary differential equations,
$$
\EQNalign{
    {1\over R} {d\over dR} ( R {d\R\over dR} )
      + ( k^2 - {m^2 \over R^2} ) \R &= 0 
      \;, \EQN ode;a \cr
    {d^2\Phi \over d\phi^2} + m^2 \Phi &= 0
      \;, \EQN ode;b \cr
    {d^2 \Z \over dz^2} - (k^2 + \lambda) \Z &= 0
      \;, \EQN ode;c \cr
}
$$
where $k$ and $m$ are separation constants; $k$ is real and can be
taken positive while $m$ is an integer due to periodicity of the
$\phi$ coordinate.  These equations~\Ep{ode} have the well-known
solutions
$$
\EQNalign{
    \R(R) &= J_m(k R)
      \;, \EQN ode soln;a \cr
    \Phi(\phi) &= e^{i m \phi}
      \;, \EQN ode soln;b \cr
    \Z(z) &= e^{\pm \Sqrt{k^2 + \lambda} \, z}
      \;, \EQN ode soln;c \cr
}
$$
where $J_m$ is the cylindrical Bessel function of order~$m$.

It is easy to see that the eigenvalue $\lambda$ must be negative
[multiply \Eq{eval} by $\psi$ and integrate by parts using Green's
theorem (\eg, Arfken 1985, eq.~1.98) to get $-\int(\nabla\psi)^2\,dV =
\lambda\int\psi^2\,dV$].
For our purpose of obtaining a basis set, we can impose the further
restriction that $\lambda \le -k^2$.  Thus we may write $\Z(z) =
e^{ihz}$, where $h$ is any real number; the eigenvalue is then
$\lambda = -(k^2 + h^2)$.  The eigenfunctions are
$$
\psi_{kmh} (R,\phi,z) =
  J_m(k R) \, e^{im\phi} \, e^{ihz}
\;.
\EQN pair;a
$$
To each eigenpotential corresponds the density
$$
\rho_{kmh} (R,\phi,z) =
  -{k^2 + h^2 \over 4\pi G}\,
  J_m(k R) \, e^{im\phi} \, e^{ihz}
\;.
\EQN pair;b
$$
By standard theorems (\eg, Courant \& Hilbert 1953, \S6.3; Arfken
1985, \S9.4) these functions form a complete, orthogonal basis for the
space of square-integrable functions, $\scriptL^2(\reals^3)$.  Note
that to obtain correct units explicitly, a factor $GM/\length$ should
be appended to $J_m(k R)$, where $\length$ and $M$ are length and mass
scales.

The standard basis of eigenfunctions \Ep{pair} is very simple, but it
is poorly suited for numerical work with density distributions that
are nearly flat, like disk galaxies.  In particular, individual basis
functions do not satisfy the natural galactic boundary condition,
$\psi\to0$ as $R^2+z^2\to\infty$.  At the very least we need basis
functions that decay both as $R\to\infty$ and as $z\to\pm\infty$.

\section{Basis functions for galactic disks}

We can derive more suitable basis sets by relaxing the conditions we
imposed in \Eqs{eval} and \Ep{sep}.  Rather than demanding
eigenfunctions, we shall specify the vertical density profile
$\zden(z)$ in advance and seek solutions of Poisson's
equation~\Ep{Poisson} with
$$
\EQNalign{
  \psi(R,\phi,z) &= \R(R) \Phi(\phi) \Z(z) \;, \EQN idea;a \cr
  \rho(R,\phi,z) &= {1\over 4\pi G} \,
            \R(R) \Phi(\phi) \zden(z) \;. \EQN idea;b \cr
}
$$
Note the difference between this and ordinary separation of variables:
here we are free to choose the form of $\zden(z)$ and we will not have
$\Z(z)\propto \zden(z)$, unless we choose $\zden(z)=e^{ihz}$.

In this case, \Eq{Poisson} again separates into three ordinary
differential equations, namely \Eqs{ode;a} and
\Ep{ode;b} together with
$$
{d^2 \Z \over dz^2} - k^2 \Z = \zden(z) \;,
\EQN z ode
$$
which replaces \Eq{ode;c}.

We are interested in bell-shaped density factors $\zden(z)$ that
resemble actual vertical density profiles of disk galaxies, and we
impose the boundary conditions that the potential factor $\Z(z)\to0$ as
$z\to\pm\infty$.  Many analytical solutions of this form can be found
for \Eq{z ode}.  Several simple and useful solutions are listed in
\Tbl{no k}.  The first row gives Green's function for \Eq{z ode};
when inserted in \Eq{idea} we obtain the well-known Bessel function
basis set for flat disks, used by Toomre (1981) for his flat disk
stability analysis.  Rows 2--4 give examples that allow us to
construct 3D basis sets for realistic disk galaxy models.

\def\exponentialpot{
  \cases{
         -{1\over k^2 - \alpha^2}
             ( e^{-\alpha\abs{z}} - {\alpha\over k} e^{-k\abs{z}} ) \;,
                 & $k\ne\alpha$ \crnorule
         -{1\over2k^2}\,({1} + k\abs{z}) \,e^{-k\abs{z}} \;,
                 & $k=\alpha$ \crnorule
        }
}
\def\gaussianpot{
        -{e^{(k/2\alpha)^2} \sqrt{\pi} \over 4\,k\,\alpha}
        \left[ e^{kz} \erfc({k\over2\alpha} + \alpha z)
        +     e^{-kz} \erfc({k\over2\alpha} - \alpha z)
        \right]
}

\def\sechpowerpot{
\eqalign{
        -{2^{q-1} \over k^2 + q\,k\,\alpha}
        & \left[ e^{q\,\alpha\, z}
          \hyg(q,{q\over 2} + {k\over {2\,\alpha }};
                 1 + {q\over 2} + {k\over {2\,\alpha }};-{e^{2\,\alpha \,z}}) 
                 \right. \crnorule
        & + \left.
          e^{-q\,\alpha\, z}
          \hyg(q,{q\over 2} + {k\over {2\,\alpha }};
                 1 + {q\over 2} + {k\over {2\,\alpha }};-{e^{-2\,\alpha \,z}}) 
        \right] \crnorule
        }
}
\widetable{no k}
\caption{Useful solutions of \Eq{z ode}}
\tenpoint

%
\thicksize=\thinsize
\ruledtable
 Density factor $\zden(z)$        & Potential factor $\Z(z)$ \CR
 $\delta(z)$                      | $e^{-k\abs{z}}$   \cr
 $e^{-\alpha\abs{z}}$             | $\exponentialpot$ \cr
 $e^{-\alpha^2 z^2}$              | $\gaussianpot$    \cr
 $\sech^q(\alpha z)$              | $\sechpowerpot$ 
\endruledtable
\endtable

The density factors $\zden(z)$ in \Tbl{no k} do not depend on the
radial index $k$.  As a result, the full 3D potential of any disk of
the form~\Ep{densep} with $\zden(z)$ drawn from \Tbl{no k} can be
obtained by a single integration over $k$, provided the Hankel
transform of the radial density profile $\Rden(R)$ is known
analytically:
$$
\psi(R,z) = \zden(z) \int_0^\infty J_0(kR)\, \Z(z)\, S(k)\, dk \,,
\EQN
$$
where $S(k)$ is the Hankel transform of $\Rden(R)$.
In this context, the solutions in 
\Tbl{no k} have been found previously (\eg, Casertano 1983, Kuijken \& Gilmore
1989, Sackett \& Sparke 1990, Cudderford 1993).
The principal contribution of this paper is to show that these
solutions, and simpler solutions given below,
can be used to construct complete 3D basis sets.

If we are willing to let $\zden(z)$ itself depend on the radial wave
number $k$ then we can find somewhat simpler solutions from which to
construct complete basis sets (\Tbl{with k}).  Several basis functions
for each $k$ are then required to represent a fixed vertical density
profile such as $\sech^2(\alpha z)$.  However, this may impose no
practical disadvantage for applications where the system evolves
and/or where only the large-scale normal modes of the model are
required.

\bigskip\bigskip\bigskip
\widetable{with k}
\caption{Simpler Solutions with $\zden(z)$ depending on $k$}
\singlespaced\tenpoint

\thicksize=\thinsize
\ruledtable
 Density factor $\zden(z)$ 
      & Potential factor $\Z(z)$\cr
$k^2\,e^{-k\abs{z}}$
      | $-{1\over2}(1 + k\abs{z})\,e^{-k\abs{z}}$ \crnorule
$k^2\,e^{-(kz)^2}$ 
      | $-{1\over4}e^{1/4} \sqrt{\pi} \,
       [ e^{kz} \erfc(1/2 + kz) + e^{-kz} \erfc(1/2 - kz) ]$ \crnorule
$k^2\,\sech{(kz)}$
      | $kz\,e^{kz} - \cosh{(kz)}\log{(1+e^{2kz})}$ \crnorule
$k^2 \, \sech^2{(kz)}$ 
      | $1 + \sinh(kz) \,\arctan[\sinh(kz)] - {\pi\over2}\cosh(kz)$  
\endruledtable
\endtable
\bigskip\bigskip

\section{New basis sets}

So far we have merely found potential-density pairs; we do not yet
have 3D basis sets.

\subsection{Construction}

To form basis sets from the potential-density pairs indicated in
\Tbl{no k} and \Tbl{with k}, we make the following observation: the 
left hand side of \Eq{z ode} is invariant under the translation
$z\to(z-h)$ for any $h$, so if we replace $\zden(z)$ by $\zden(z-h)$
and $\Z(z)$ by $\Z(z-h)$ then we have another potential-density pair.
This corresponds to shifting the original configuration up a distance
$h$.  As will be shown in the next subsection, if we let $h$ vary from
$-\infty$ to $\infty$ then the functions
$$
\EQNalign{
\psi_{kmh} (R,\phi,z) & =   J_m(k R) \, e^{im\phi} \, \Z(z-h) \;,
	     \EQN new pair;a \cr
\rho_{kmh} (R,\phi,z) &=
  {1\over 4\pi G}\, J_m(k R) \, e^{im\phi} \, \zden(z-h) \;,
  \EQN new pair;b \cr
}
$$
form an $\scriptL^2$-complete basis, where $k\in\reals^{>0}$,
$m\in\integers$, and $\Z(z)$ and $\zden(z)$ are taken from any row in
\Tbl{no k} or \Tbl{with k}.  Representing a given model amounts to
weighting and stacking the functions~\Ep{new pair} along the $z$-axis.

\subsection{Completeness}

It is sufficient to show that any member of the standard
basis~\Ep{pair} can be represented.  Since our sets differ from the
standard basis only in their vertical factors $\zden(z)$, it is enough
to prove that for any $h\in\reals$ there exists a function $w_h(h')$
such that
$$
\int_{-\infty}^\infty w_h(h')\,\zden(z-h') \,dh'
= e^{ihz} \;.
\EQN Fredholm
$$
This Fredholm integral equation of the first kind can be solved
explicitly analytically (\cf\ Titchmarsh 1937).  Taking the Fourier
transform of \Eq{Fredholm} and using the convolution theorem, we have
$$
\ftwh(u)\,\ftzden({u})
= {1\over\sqrt{2\pi}} \int_{-\infty}^\infty e^{ihz} \,e^{iuz} \,dz
= \delta(u+h) \;,
\EQN
$$
where $\ftwh(u)$ and $\ftzden(u)$ denote the Fourier transforms of $w_h(h')$ and
$\zden(h')$ respectively, and $\delta$ is the Dirac delta
distribution.  It follows immediately that
$$
w_h(h') = {1\over\sqrt{2\pi}} \int_{-\infty}^\infty 
       {\delta(u+h) \over \ftzden({u})} \,e^{-iuh'} \,du
     = {1\over\sqrt{2\pi}} {e^{ihh'} \over \ftzden(-{h})}  \;.
\EQN wh
$$
This formula is meaningful provided $\zden(z)$ has a Fourier transform
that is nowhere zero.  \Tbl{ft} gives the Fourier transforms of all
the functions in \Tbl{no k} and \Tbl{with k} [except for arbitrary
powers of $\sech(\alpha z)$; for that transform see Oberhettinger 1973,
\S{I.7.211}].
Each of these is strictly positive,
so our new basis sets are complete in $\scriptL^2(\reals^3)$.

\midtable{ft}
\caption{Fourier transforms}
\singlespaced\tenpoint

\thicksize=\thinsize
\ruledtable
 Density factor $\zden(z)$ 
      & $\ftzden(u)={1\over\sqrt{2\pi}}
        \int_{-\infty}^\infty \zden(z)\,e^{iuz}\,dz$\cr
$\delta(z)$
      | ${1\over\sqrt{2\pi}}$ \crnorule                  
$e^{-\alpha\abs{z}}$
      | $\big({2\over\pi}\big)^{1/2}{\alpha \over \alpha^2 + u^2}$ \crnorule     
$e^{-(\alpha z)^2}$ 
      | ${1\over\sqrt{2} \alpha}\, e^{-u^2/4\alpha^2}$ \crnorule   
$\sech{(\alpha z)}$
      | ${1\over2}\sech{(\pi u/2\alpha)}$ \crnorule           
$\sech^2{(\alpha z)}$ 
      | ${\pi\over4\alpha}\,u\,\csch{(\pi u/2\alpha)}$             
\endruledtable
\endtable


\subsection{Expansion coefficients}

The inner product of two potential-density pairs is defined by minus
the interaction potential energy of the two densities $\rho$ and
$\rho'$,
$$
\eqalign{
\inprod{\rho\,|\,\psi'}
  & \equiv - {\length\over GM^2} \int \rho^* \psi' \,dV  \cr
  & = -{\length\over4\pi G^2M^2} \int (\nabla^2 \psi^*) \psi' \,dV \;. \cr
}
\EQN innerprod
$$
Here, the asterisk denotes complex conjugation and the factor
$\length/GM^2$ 
makes the inner product dimensionless.  All functions are assumed to
live in $\scriptL^2(\reals^3)$.  In cylindrical coordinates,
$$
\inprod{\rho\,|\,\psi'} = -\int_{-\infty}^\infty \int_0^{2\pi}\!\int_0^\infty
	 \rho^* \psi'\,R\,dR \,d\phi \,dz \,,
\EQN
$$
where we have set $G=M=\length=1$ for convenience.

A basis is a linearly independent set that spans the full vector space
of potential-density pairs.  Thus for any basis set
$\{(\psi_j,\rho_j):\nabla^2\psi_j=4\pi\rho_j\}$ and any given density
$\rho\in\scriptL^2(\reals^3)$ there is a set of complex coefficients
$\{c_j\}$ such that $\rho = \sum_j c_j \rho_j$.
It follows immediately that $\inprod{\rho\,|\,\psi_{j'}}=\sum_j c_j
\inprod{\rho_j\,|\,\psi_{j'}}$, so defining
$$
{\cal M}_{j{j'}}\equiv\inprod{\rho_j\,|\,\psi_{j'}} \,,
\EQN Mij
$$
we have
$$
c_j = \sum_{j'} {\cal M}_{j{j'}}^{-1} \inprod{\rho\,|\,\psi_{j'}} \,,
\EQN
$$
where ${\cal M}_{j{j'}}^{-1}$ is the $jj'$ element of matrix inverse
of ${\cal M}$.  The potential generated by $\rho$ is $\psi = \sum_j
c_j \psi_j$.  A basis is said to be {\it biorthogonal\/} if ${\cal
M}_{j{j'}}$ is diagonal, \ie, $\inprod{\rho_j\,|\,\psi_{j'}} \neq 0$
if and only if $j=j'$, and {\it biorthonormal\/} if ${\cal M}_{j{j'}}$
is the identity matrix $\delta_{jj'}$.  With a biorthonormal basis,
the dimensionless expansion coefficients are simply $c_j =
\inprod{\rho\,|\,\psi_j} = \inprod{\rho_j\,|\,\psi}$.  Any finite 
subset of a basis can be made biorthonormal with the Gram-Schmidt
algorithm (\eg, Arfken 1985).  Biorthogonality is clearly preferable,
but without it the only extra effort required is the one-time
evaluation and inversion of the matrix ${\cal M}_{jj'}$~\Ep{Mij}, as
emphasized by Saha (1993).

In the present context, where the abstract index $j$ represents the
triplet $(k,m,h)$ with $k$ and $h$ real,
the sum over $j$ refers to a discrete sum and two integrals,
$$
\EQNalign{
\rho &= \sum_{m=-\infty}^\infty \int_{-\infty}^\infty \int_0^\infty
	c_{kmh}\,\rho_{kmh}\,dk\,dh \,, \EQN PD;a \cr
\psi &= \sum_{m=-\infty}^\infty \int_{-\infty}^\infty \int_0^\infty
	c_{kmh}\,\psi_{kmh}\,dk\,dh \,. \EQN PD;b \cr
}
$$
The matrix ${\cal M}_{j{j'}}$ refers to
$$
{\cal M}_{kmh\,k'm'h'} = \inprod{\rho_{kmh}\,|\,\psi_{k'm'h'}} \,,
\EQN
$$
and biorthonormality means strictly that
$$
{\cal M}_{kmh\,k'm'h'} = \delta(k-k')\, \delta_{mm'} \,\delta(h-h') \;.
\EQN biorthonormal
$$
Using the Bessel closure relation,
$$
\int_0^\infty J_m(kR) J_m(k'R) \,R\,dR = {1\over k}\,\delta(k-k') \,,
\EQN
$$
(\eg, Arfken 1985, Eq.~11.59) and the exponential identity,
$$
\int_0^{2\pi} e^{-im\phi} e^{im'\phi} \,d\phi = 2\pi\delta_{mm'} \,,
\EQN
$$
we find for the present class of basis sets~\Ep{new pair},
$$
{\cal M}_{kmh\,k'm'h'} = \delta(k-k')\, \delta_{mm'} \,{\cal M}_{khh'}\,,
\EQN
$$
where
$$
{\cal M}_{khh'} \equiv -{1\over2k} \int_{-\infty}^\infty \zden^*(z-h) \Z(z-h') \,dz \;.
\EQN
$$
The expansion coefficients are therefore
$$
\EQNalign{
c_{kmh} &= \sum_{m'=-\infty}^\infty \int_{-\infty}^\infty \int_0^\infty
		{\cal M}_{kmh\,k'm'h'}^{-1} \inprod{\rho\,|\,\psi_{k'm'h'}}
		\,dk'\,dh' \cr
	&= \int_{-\infty}^\infty {\cal M}_{khh'}^{-1} \inprod{\rho\,|\,\psi_{kmh'}}
		\,dh' \,, \EQN c kmh \cr
}
$$
where ${\cal M}^{-1}$ is the matrix inverse of ${\cal M}$.  To find
${\cal M}_{khh'}^{-1}$ we must solve the integral equation
$$
\int_{-\infty}^\infty {\cal M}_{khh''}^{-1} {\cal M}_{kh''h'} \,dh'' = \delta(h-h') \,.
\EQN inteq
$$
We discuss how to solve this equation numerically in the next section.

\section{Discretization}

We have given several crucial formulae that involve integrals over the
radial and vertical basis indeces, $k$ and $h$ [\Ep{PD}, \Ep{c kmh},
\Ep{inteq}].  Since our goal is to use as few basis functions as possible
to obtain a given accuracy, these integrations must be reduced to
computations with a small, finite number of $k$ and $h$ values.

\subsection{Numerical quadrature}

The continuous indeces $k$ and $h$ can be replaced with discrete
indeces $n$ and $\ell$ via prescriptions of the form
$$
\EQNalign{
\int_0^\infty f(k)\,dk &\longrightarrow \sum_{n=0}^N w^k_n f(k_n) \,,
  \EQN int2sum;a \cr
\int_{-\infty}^\infty g(h)\,dh &\longrightarrow \sum_{\ell=0}^L w^h_\ell g(h_\ell) \,,
  \EQN int2sum;b \cr
}
$$
where the sums approximate the integrals to some given order in $1/N$
and $1/L$.  The simplest approach is to choose a set of evenly spaced
abscissae, $k_n = n\Delta k$ and $h_\ell = (\ell-L/2)\Delta h$, and to
use a classical formula such as Simpson's rule to define the weights
(\eg, Abramowitz \& Stegun 1972, \S25.4.6; Press \etal\ 1992,
Eq.~4.1.13).  In the case of the axisymmetric ($m=0$) integrals over
$k$ in \Eq{PD}, we can do better by employing a Gaussian quadrature
rule (\eg, Press \etal\ 1992, \S4.5) because we know the initial
radial profile.

\subsection{Solving the integral equation for ${\cal M}^{-1}$}

In any application, our first numerical task is to solve \Eq{inteq}
for ${\cal M}^{-1}_{khh'}$.  To reduce notational complexity, let us
define
$$
{\cal A}^n_{\ell\ell'} \equiv {\cal M}_{k_n h_\ell h_{\ell'}} \,,
\qquad
{\cal B}^n_{\ell\ell'} \equiv {\cal M}^{-1}_{k_n h_\ell h_{\ell'}} \,.
\EQN
$$
We wish to solve for the matrices ${\cal B}^n$, $n=0,\ldots,N$.  Using
\Ep{int2sum}, \Eq{inteq} becomes
$$
\EQNalign{
\sum_{\ell''=0}^L w^h_{\ell''} {\cal B}^n_{\ell\ell''} {\cal A}^n_{\ell''\ell'} 
  &= 0 \,, \qquad \ell\ne\ell' \,,
  \EQN mateq;a \cr
w^h_{\ell}\sum_{\ell''=0}^L w^h_{\ell''} {\cal B}^n_{\ell\ell''} {\cal A}^n_{\ell''\ell} 
  &= 1\,,
  \EQN mateq;b \cr
}
$$
where the second line is obtained by integrating \Eq{inteq} in a small
interval about $h_\ell$ for $\ell=\ell'$.  Putting these two pieces together
we have
$$
\sum_{\ell''=0}^L w^h_{\ell''} {\cal B}^n_{\ell\ell''} {\cal A}^n_{\ell''\ell'} 
  = {1\over w^h_{\ell}} \delta_{\ell\ell'} \,.
\EQN
$$
Thus ${\cal B}^n$ is the ordinary matrix inverse of ${\cal A}^n{\cal W}$,
where ${\cal W}={\rm diag}(w^h_0,\ldots,w^h_L)$.
Each matrix ${\cal B}^n$
needs to be computed only once for a given basis set, so the work
involved is always negligible.

\subsection{Discrete expansion formulae}

We can now rewrite our formulae for the expansion coefficients and
expanded potential in a completely discrete manner.  \Eq{c kmh}
becomes
$$
c_{nm\ell} = \sum_{\ell'} w^h_{\ell'} \, {\cal B}^n_{\ell\ell'}
  \inprod{\rho\,|\,\psi_{nm\ell'}} \,.
\EQN dis1
$$
\Eq{PD;b} becomes
$$
\psi = \sum_m \sum_\ell w^h_\ell \sum_n w^k_n \, c_{nm\ell} \, \psi_{nm\ell} \,,
\EQN dis2
$$
and a similar formula replaces \Eq{PD;a}.  Note that in the case of a
distribution of point particles, $\rho=\sum_i M_i\, \delta(\x-\x_i)$, the
inner product in \Ep{dis1} also becomes a discrete sum,
$$
\inprod{\rho\,|\,\psi_{nm\ell}} = 
  -\sum_i M_i\, \psi_{nm\ell}(\x_i) \,.
\EQN dis3
$$

\section{Discussion}

A similar approach can be used to derive basis sets for 3D elliptic
disks.  In fact, all we need to do is replace the factor
$J_m(kr)\,e^{im\phi}$ of our basis sets with the eigenfunctions of the
Laplacian in elliptic coordinates (Mathieu functions, \eg, Abramowitz
\& Stegun 1972, chapter 20).  Such basis sets should be useful for
stability studies of realistically thickenned versions of flat
elliptic disks (\eg, Evans \& de Zeeuw 1992).

For stability work, it is essential to be able to represent the {\it
normal modes\/} of the system with a small number of terms of the
chosen basis expansion.  There is no guarantee that a basis selected
for its ability to represent the unperturbed model will be ideal for
representing its normal modes.  It is prudent, therefore, to repeat
stability calculations with several different basis sets.  Robijn \&
Earn (1996) discuss alternatives to the present class.

Both in stability analyses and $N$-body experiments, only the basis
potentials are used for the principal computations: in \Eqs{dis1},
\Ep{dis2} and \Ep{dis3} the basis densities do not appear.
For this reason, it is unfortunate that in \Tbls{no k} and \Tblp{with
k} the potential functions are more complicated than the associated
densities.  Nevertheless, they are easy to implement because the
special functions they involve are available in all standard numerical
libraries.  Moreover, all the potentials and densities discussed in
this paper are separable so they can be interpolated efficiently and
accurately from one-dimensional tables.  The basis sets provided here
should, therefore, be useful for stability studies and modeling of
disk galaxies.

\bigskip
\medskip

I am grateful to Tim de Zeeuw, Ed Doolittle, and Alar Toomre for
helpful discussions and correspondence, and to Frank Robijn both for
discussions and detailed comments on a preliminary version of the
manuscript.  I also thank the referee, Agris Kalnajs, for useful
comments.  I was supported by a Lady Davis Postdoctoral Fellowship.

\vfill\eject
\nosechead{References}

\begingroup
\tenpoint
\parindent=0pt\everypar{\hangindent 1cm}
\nobreak\vskip 6pt
\def\papertitle#1{\relax}
\def\booktitle#1{{\sl#1\/}}
%


Abramowitz M, Stegun IA, 1972, 
\booktitle{Handbook of Mathematical Functions,}
(Dover: New York)

Arfken G, 1985,
\booktitle{Mathematical Methods for Physicists,}
Third Edition (\AP)

Binney JJ, Tremaine S, 1987, 
\booktitle{Galactic Dynamics} (\PUP)

Casertano S,
\journal \papertitle{The rotation curve of the edge-on spiral galaxy NGC5907:
disc and halo masses}
\MNRAS;203,735--747(1983)

Clutton-Brock M, 
\journal \papertitle{The Gravitational Field of Flat Galaxies}
\ApSS;16,101--119(1972)

Clutton-Brock M, 
\journal \papertitle{The Gravitational Field of Three-Dimensional Galaxies}
\ApSS;23,55--69(1973)

Courant R, Hilbert D, 1953,
\booktitle{Methods of Mathematical Physics,}
Vol.~1 (Interscience: New York)

Cudderford P,
\journal \papertitle{On the potentials of galactic discs}
\MNRAS;262,1076--1086(1993)

Earn DJD, 1995,
in \booktitle{Unsolved problems of the Milky Way}, Proc.\ IAU Symp.\ 169,
edited by L.\ Blitz, in press

Earn DJD, Sellwood JA,
\journal \papertitle{A high quality $N$-body method for stability studies of galaxies}
\ApJ;451,533--541(1995)


Evans NW, de Zeeuw PT,
\journal \papertitle{Potential-density pairs for flat galaxies}
\MNRAS;257,152--176(1992)

Freeman KC,
\journal \papertitle{On the disks of spiral and S0 galaxies}
\ApJ;160,811--830(1970)


Hernquist L, Ostriker JP,
\journal \papertitle{A Self-Consistent Field Method for Galactic Dynamics}
\ApJ;386,375--397(1992)

Kalnajs AJ,
\journal \papertitle{The Equilibria and Oscillations 
of Family of Uniformly Rotating Stellar Disks}
\ApJ;175,63--76(1972)

Kalnajs AJ,
\journal \papertitle{Dynamics of flat galaxies. 
II.~Biorthonormal surface density-potential pairs for finite disks}
\ApJ;205,745--750(1976)

Kalnajs AJ,
\journal \papertitle{Dynamics of Flat Galaxies. IV.~The Integral
Equation for Normal Modes in Matrix Form}
\ApJ;212,637--644(1977)

Kuijken K, Gilmore G,
\journal \papertitle{}
\MNRAS;239,571--603(1989)

Morse PM, Feshbach H, 1953,
\booktitle{Methods of theoretical physics,}
(McGraw-Hill: New York)

Oberhettinger F, 1973,
\booktitle{Fourier Transforms of Distributions and Their Inverses:
a collection of tables}
(\AP: New York)

Press WH, Teukolsky SA, Vetterling WT, Flannery BP, 1992,
\booktitle{Numerical Recipes in C,}
Second Edition (\CUP)

Qian EE,
\journal \papertitle{Potential-density pairs for flat discs}
\MNRAS;257,581--592(1992)

Qian EE,
\journal \papertitle{Biorthogonal potential-density sets for flat discs}
\MNRAS;263,394--402(1993)

Robijn FHA, Earn DJD,
\journal \papertitle{Potential-density basis sets in axisymmetric coordinates}
\MNRAS;xx,in press(1996)

Sackett P, Sparke L,
\journal \papertitle{The dark halo of the polar-ring galaxy NGC 4650A}
\ApJ;361,408--418(1990)

Saha P,
\journal \papertitle{Designer basis functions for potentials in galactic
dynamics}
\MNRAS;262,1062--1064(1993)

Spitzer L,
\journal \papertitle{The dynamics of the interstellar medium.
III.\ Galactic distribution}
\ApJ;95,329--344(1942)

Syer D,
\journal
\MNRAS;276,1009(1995)

Titchmarsh EC, 1937,
\booktitle{Introduction to the Theory of Fourier Integrals}
(Chelsea: New York)


Toomre A, 1981,
\papertitle{What Amplifies the Spirals?}
in \booktitle{The Structure and Evolution of Normal Galaxies,}
edited by S.M.~Fall and D.~Lynden-Bell (\CUP) pp.~111--136

van der Kruit PC,
\journal \papertitle{The three-dimensional distribution of light and mass 
in disks of spiral galaxies}
\AAp;192,117-127(1988)

Zhao HS,
\journal \papertitle{Analytical models for galactic nuclei}
\MNRAS;xx,in press(1996)

\endgroup


\bye